\documentclass[conference]{IEEEtran}
\IEEEoverridecommandlockouts
\usepackage{cite}
\usepackage{amsmath,amssymb,amsfonts}
\usepackage{algorithmic}
\usepackage{graphicx}
\usepackage{textcomp}
\usepackage{multirow}
\usepackage{amsmath,array,graphicx}
\usepackage{kantlipsum}
\usepackage{blindtext}
\usepackage{booktabs}
\usepackage[table,xcdraw]{xcolor}
\usepackage{dblfloatfix}% http://ctan.org/pkg/dblfloatfix
\usepackage{lipsum}% http://ctan.org/pkg/lipsum
\usepackage{xcolor}
\def\BibTeX{{\rm B\kern-.05em{\sc i\kern-.025em b}\kern-.08em
    T\kern-.1667em\lower.7ex\hbox{E}\kern-.125emX}}
\usepackage{amsthm}
\theoremstyle{definition}
\newtheorem{definition}{Definition}[section]
\theoremstyle{remark}

\begin{document}

\title{Finding Small and Large $k$-Clique Instances on a Quantum Computer\\
%\thanks{This work is funded by QLEAP}
}

\author{\IEEEauthorblockN{Sara Ayman Metwalli}
\IEEEauthorblockA{Keio Quantum Computing Center} \\
\textit{Keio University}\\
Fujisawa, Japan \\
{sara@sfc.wide.ad.jp}
\and
\IEEEauthorblockN{Fran{\c c}ois Le Gall}
\IEEEauthorblockA{Graduate School of Mathematics} \\
\textit{Nagoya University}\\
Nagoya, Japan \\
{legall@math.nagoya-u.ac.jp}
\and
\IEEEauthorblockN{Rodney Van Meter}
\IEEEauthorblockA{Keio Quantum Computing Center} \\
\textit{Keio University}\\
Fujisawa, Japan \\
{rdv@sfc.wide.ad.jp}
}

\maketitle

\begin{abstract}
Algorithms for triangle-finding, the smallest nontrivial instance of the $k$-clique problem, have been proposed for quantum computers. Still, those algorithms assume the use of fixed access time quantum RAM (QRAM). We present a practical gate-based approach to both the triangle-finding problem and its NP-hard k-clique generalization. We examine both constant factors for near-term implementation on a Noisy Intermediate Scale Quantum computer (NISQ) device, and the scaling of the problem to evaluate long-term use of quantum computers. We compare the time complexity and circuit practicality of the theoretical approach and actual implementation. We propose and apply two different strategies to the $k$-clique problem, examining the circuit size of Qiskit implementations. We analyze our implementations by simulating triangle finding with various error models, observing the effect on damping the amplitude of the correct answer, and compare to execution on six real IBMQ machines. Finally, we estimate the date when the methods proposed can run effectively on an actual device based on IBM’s quantum volume exponential growth forecast and the results of our error analysis.
\end{abstract}

\begin{IEEEkeywords}
Quantum computing, clique, graph, Grover's algorithm
\end{IEEEkeywords}

\section{Introduction}

A \emph{clique} is defined as a complete subgraph over a subset of vertices in an undirected graph. Several computational problems address finding cliques in a given graph. These problems vary based on what information about the clique needs to be found. One such is the $k$-clique problem, which answers the question, ``Given an undirected graph and a positive integer $k$, does a clique with size $k$ exist?'' 
The $k$-clique problem is NP-Complete for large values of $k$, as shown by Karp \cite{karp1972reducibility} and Cook~\cite{cook1971complexity}. 
Probably one of the most studied version of the $k$-clique problem is the triangle finding problem (the 3-clique problem), which has been addressed both classically \cite{castellanos2002triangle}, \cite{williams2014finding} and quantumly \cite{le2014improved},\cite{magniez2007quantum}. The best known classical algorithm has time complexity O($n^{2.38}$) while the best known quantum algorithm has time complexity O($n^{1.5}$), where n is the number of nodes in the graph and N is the sizeof the search space ($n^2$).
Several quantum algorithms have also been proposed for the $k$-clique problem with $k>3$ \cite{magniez2007quantum}, \cite{childs2000finding},  \cite{childs2003quantum}. 

In this paper, we present several implementations based on Grover's algorithm~\cite{grover1996fast}.  The asymptotic behavior of the algorithm tells us that quantum computers will offer better scaling than classical computers for a broad range of problems in the long run. However, we must also assess the constant factors, especially when considering near-term implementation on a NISQ device (section~\ref{imp}). We use the algorithm to solve the $k$-clique problem using Dicke or W states to limit the search space, studying the trade-off against circuit size. We address the theoretical complexity of the above algorithms, which assume the existence of constant access time QRAMs, whereas the best proposed approach would be $O(\log N)$ access time \cite{giovannetti2008quantum}. Moreover, implementation of even logarithmic access time memory is not yet possible. Instead, gate-based representations of graphs are necessary, with cost that exceeds the polynomial gains promised by using Grover for triangle finding (section~\ref{res}). Our work aims to decrease the gap between theory and implementation by presenting a robust implementation of the $k$-clique problem in general, regaining the quantum advantage for larger problems.

We implement our proposed scheme using Python and the Quantum Science Kit (Qiskit)\footnote{The Qiskit version used in this work is '0.15.0'} developed by IBM \cite{aleksandrowicz2019qiskit}. Then based on our data, we predict when the smallest instance of the $k$-clique problem (triangle finding) can be executed with minimal error on a real quantum computer (section~\ref{dis}). Finally, we conclude the paper with some discussions and future directions (section~\ref{con}).

%\footnotetext{The Qiskit version used in this work is '0.15.0'}
\section{Background}
\label{bg}

In this section, we will lay out some background knowledge on Grover's Search algorithm, the $k$-clique problem, and the Dicke states.

\subsection{Grover's Search Algorithm}
\begin{figure}[htbp]
\centerline{\includegraphics[scale=0.35]{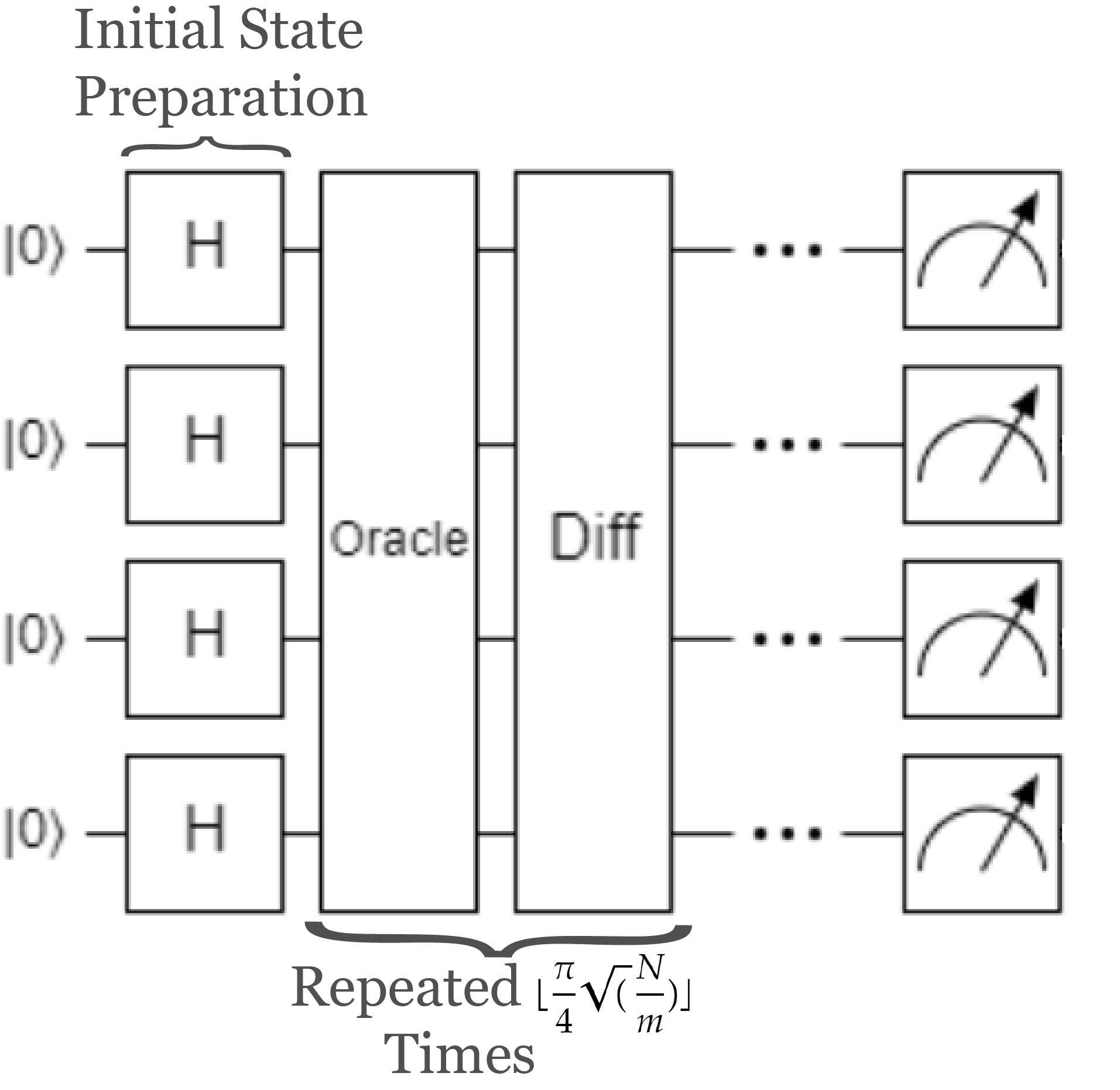}}
\caption{An Overview of Grover's algorithm's Steps}
\label{grov}
\end{figure}

Grover's Search Algorithm answers the question ``Given a function $f(x)$, what values of $x$ cause $f(x)$ to evaluate to {\bf True}?''. The algorithm presents a framework for tackling the search problem in an unsorted database with complexity O($\sqrt{N}$). It mainly consists of three sections, state preparation, the oracle, and the diffusion operator, which can be seen in Fig.~\ref{grov}.\footnote{Circuit illustration is created using Quirk~\cite{quirk}}

The algorithmic steps of Grover's search are:
\begin{enumerate}
    \item Prepare the input in a symmetric-superposition state.
    \item Apply Grover's Oracle to the prepared state.
    \item Apply the diffusion operator to the oracle's results.
    \item Iterate over step 2 and step 3 until the answer is reached.
\end{enumerate}

The first step of Grover's algorithm is preparing the initial state. In the simplest version, the initial state is prepared in an equal superposition over the entire Hilbert space. That is done by applying the Hadamard gate to all input qubits. In this paper, however, we use another approach to create entangled symmetric states as an input to the oracle (Dicke states) to decrease the size of the search space.

After the state preparation comes the oracle. The oracle is a black box function that inverts the answer by flipping its sign. Following that the diffusion operator will magnify the amplitude of the correct state while damping the amplitude of other states until the amplitude of the answer is significantly larger than the rest of the states. 
%Combined with the diffusion operator, it magnifies the state's amplitude on the following iterations until the amplitude of the answer is significantly larger than the rest of the input states.

\begin{figure}[htbp]
\centerline{\includegraphics[scale=1]{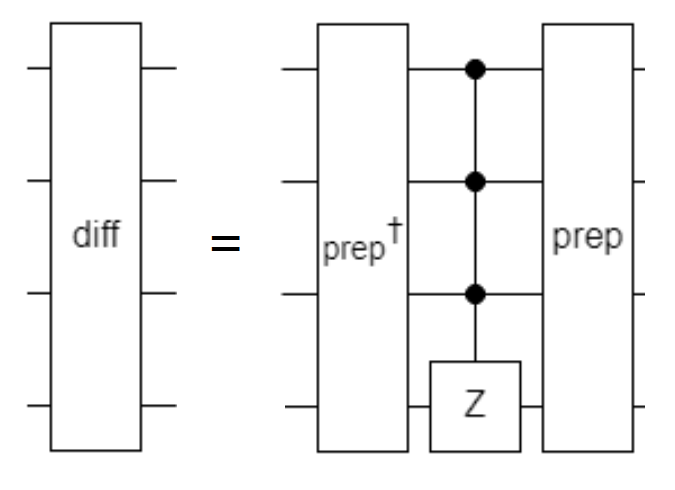}}
\caption{A General Diffusion Operator}
\label{diff}
\end{figure}

The diffusion operator is formed by: the inverse of state preparation, $C^{\otimes n}Z$ gate, state preparation, as shown in Fig.~\ref{diff}. The $C^{\otimes n}Z$ gate cost is $2n-3$ gates, divided into 1 CZ gate and $2n-4$ CCX gates.

The answer's amplitude will grow to a maximum and then decline after the optimal number of iterations {\tt opt\_iter} cyclically. Therefore, we need to measure the answer at the right time, which will lead to the first high amplitude of the answer. The optimal number of times steps 2 and 3 are repeated depends on two factors, the size of the search space $N$ and the number of answers for our search query $m$ (how many cliques in the graph) following Eq.~\ref{iter}~\cite{boyer1998tight}. For example, in Fig.~\ref{grov}, the search space is the entire Hilbert space; in this case (4 qubits), it is {$2^4$} cases 0000, 0001,....., 1111. Hence, assuming we have a question with only one answer, we will have {\tt opt\_iter} of 3.

\begin{equation}
opt\_iter =  \left\lfloor\frac{\pi}{4}\sqrt{\frac{N}{m}}\right\rfloor 
\label{iter}
\end{equation}

\subsection{Dicke States}
\label{state prep}

A Dicke state $|D_{k}^n\rangle$~\cite{PhysRev.93.99} is a fully symmetric entangled state over the $n$-qubit Hilbert space with Hamming weight $k$. For example, given a Hilbert space of 4 qubits, the Dicke state $|D_{3}^4\rangle$ will be the superposition of  $\frac{1}{2} \left( |1110\rangle + |1101\rangle + |1011\rangle + |0111\rangle \right)$ as defined in~\ref{dd}. The number of basis states with $k$ Hamming weight in a Hilbert space of $n$ qubits is \(\binom{n}{k}\).\\
%The total number of states with $k$ Hamming weight in a Hilbert space of $n$ qubits is \(\binom{n}{k}\).\\

\theoremstyle{definition}
\begin{definition}{}
Dicke state $|D_{k}^n\rangle$ is an entangled superposition of all $n$-states $|s\rangle$ with Hamming weight (hw) $k$:
\begin{equation}
\left|D_{k}^{n}\right\rangle=\left(\begin{array}{l}
{n} \\
{k}
\end{array}\right)^{-\frac{1}{2}}\sum_{s \in\{0,1\}^{n}s.t.\text {hw}(s)=k}|s\rangle.
\end{equation}
\label{dd}
\end{definition}

Dicke states represent an essential class of entangled quantum states for their applications in quantum game theory~\cite{_zdemir_2007}, quantum networking~\cite{Prevedel_2009} and quantum meteorology~\cite{T_th_2012}.
Dicke states can be implemented in several different ways; we followed the approach proposed in~\cite{bartschi2019deterministic} to prepare our Dicke states deterministically. The proposed method computes the Dicke state for any Hamming weight $k$ and $n$ qubits with $O(kn)$ gates and $O(n)$ depth~\cite{bartschi2019deterministic}. 

\subsubsection{W States}
W states are a class of entangled quantum states that are a special case of the Dicke State. W state is a Dicke state with Hamming weight 1, as shown by formula~\ref{w}. The implementation of the W-state preparation we used in this work is the algorithm proposed in~\cite{cruz2019efficient}. Since the W state is a special case of the Dicke state, the method we chose to implement the Dicke state can be used to implement a W state. However, the approach we used to implement the W state is more efficient in terms of circuit size and depth than the general Dicke state method, but it can't be extended to implement an arbitrary $k$ Hamming weight Dicke state. Hence, we refer to them as two different approaches to state preparation, and they are used to reduce the iterations needed, which will be discussed in section~\ref{time analysis}.

\begin{equation}
|W\rangle ={\frac  {1}{{\sqrt  {n}}}}(|100...0\rangle + ... + |01...0\rangle + |00...01\rangle )
\label{w}
\end{equation}

We must mention that using W states as our state preparation approach works only for clique size $k = n-1$; otherwise, W states cannot be used, and Dicke states have to be used instead.

\subsection{The $k$-clique Problem}
\begin{figure}[htbp]
\centerline{\includegraphics[scale=1.5]{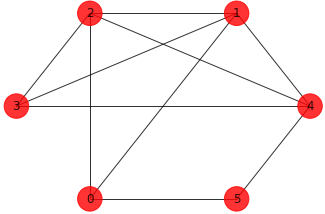}}
\caption{6-node graph with 4-clique on nodes 1, 2, 3 and 4. The output of Grover's Algorithm will be $|011110\rangle$, with 1 for every node in the clique and 0 otherwise.}
\label{graph}
\end{figure}

Given an undirected graph (G), if there exists a subset of $k$ vertices that are connected to form a complete graph, then it is said that G contains a $k$-clique — for example, Fig.~\ref{graph} represents a graph of 6 vertices, which includes a 4-clique between vertices 1, 2, 3, and 4. 

The $k$-clique problem asks us to determine if the input graph~G contains a $k$-clique, and if it does, output the vertices forming the clique \cite{valiente2013algorithms}. A popular variant of this problem only asks us to determine if G contains a $k$-clique \cite{pelillo2001encyclopedia}.
(Another adaptation of this question, which we will not consider in this paper, asks us to list all cliques of size $k$ \cite{chiba1985arboricity}.) 
Classically, several algorithms can find a clique of size $k$ in any graph with efficient complexity $O(n^k)$~\cite{vassilevska2009efficient},~\cite{regneri2007finding},~\cite{bourjolly2000heuristics}. Nevertheless, these problems become NP-complete when $k$ is large~\cite{karp1972reducibility},~\cite{cook1971complexity},~\cite{rossman2014monotone}.

%and the algorithm attempts to determine if G contains a $k$-clique, and if it does, output the vertices forming the clique. That scheme is called the $k$-clique problem~\cite{valiente2013algorithms}. Another variation of this problem is the clique decision problem. In that problem, the algorithm returns True if a $k$-clique is found and False otherwise~\cite{pelillo2001encyclopedia}.\\
Clique-finding algorithms have many practical applications. One of the main fields they can be used in is chemistry, to find chemicals matching a specific structure~\cite{rhodes2003clip}, to model molecular docking, and to find the binding sites of chemical reactions~\cite{kuhl1984combinatorial}. They can also be applied to find similar structures within different molecules~\cite{national1995mathematical}. Another field for the clique-finding algorithms is automatic test pattern generation. Finding cliques helps to confine the size of the test sets~\cite{day1986computational}. The clique-finding problems are also used for Proof-of-Work (PoW) in cryptocurrencies~\cite{lin_wang_yung_2016}. Finally, in bioinformatics, clique-finding algorithms are used to infer evolutionary trees, predict protein structures~\cite{samudrala1998graph}, and find interacting clusters of proteins~\cite{spirin2003protein}.
\section{Implementation}
\label{imp}

\begin{figure}[htbp]
\centerline{\includegraphics[scale=1.5]{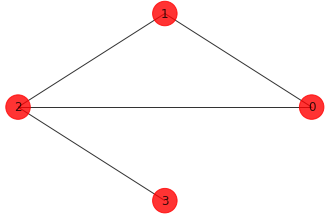}}
\caption{4-node graph containing a triangle (3-clique) on nodes 0, 1 and 2. This graph is used in Tables~\ref{tab:size-depth},\ref{tab:not-count}}
\label{exp_graph}
\end{figure}

Efficient execution of Grover is a two-fold problem: reducing the number of iterations (see~\ref{state prep} and~\ref{time analysis}), and finding a practical implementation of each iteration. 
In this section, we present two approaches to implementing the oracle circuit; we will call them the checking-oracle and the incremental-oracle, respectively. The remainder of this section will discuss both implementations in detail, starting with the checking-oracle.
Although either implementation can be used to find any $k$-clique in any given undirected graph, while explaining how both implementations work, we will consider the simplest case possible, which is a 3-clique problem (finding a triangle). In all explanations, the graph in Fig.~\ref{exp_graph} will be used.

\subsection{State Preparation}
State preparation is the first step of the implementation. Usually, when implementing Grover's algorithm, the states are prepared in an equal superposition of the whole Hilbert space using the Hadamard Gate (H gate). 
Initializing into full superposition needs only $n$ H gates and time complexity $O(1)$ since all H gates can be run simultaneously, but unnecessarily searches the full Hilbert space of all possible subsets of nodes from 0 to $n$. 

However, let's consider the case represented in Fig.~\ref{exp_graph}. If we wish to search for a 3-clique, then it makes no sense to look for a subgraph with one, two, or even four nodes. Instead, we should consider only subgraphs with $k$ nodes, and then assess whether the induced subgraph contains \(\binom{k}{2}\) edges (the number of edges in a complete graph of $k$ vertices) -- three edges in case of a triangle. Searching over a limited space should be faster. However, it will cause a significant increase in the state preparation gate count.

\begin{figure}[htbp]
\centerline{\includegraphics[scale=0.2]{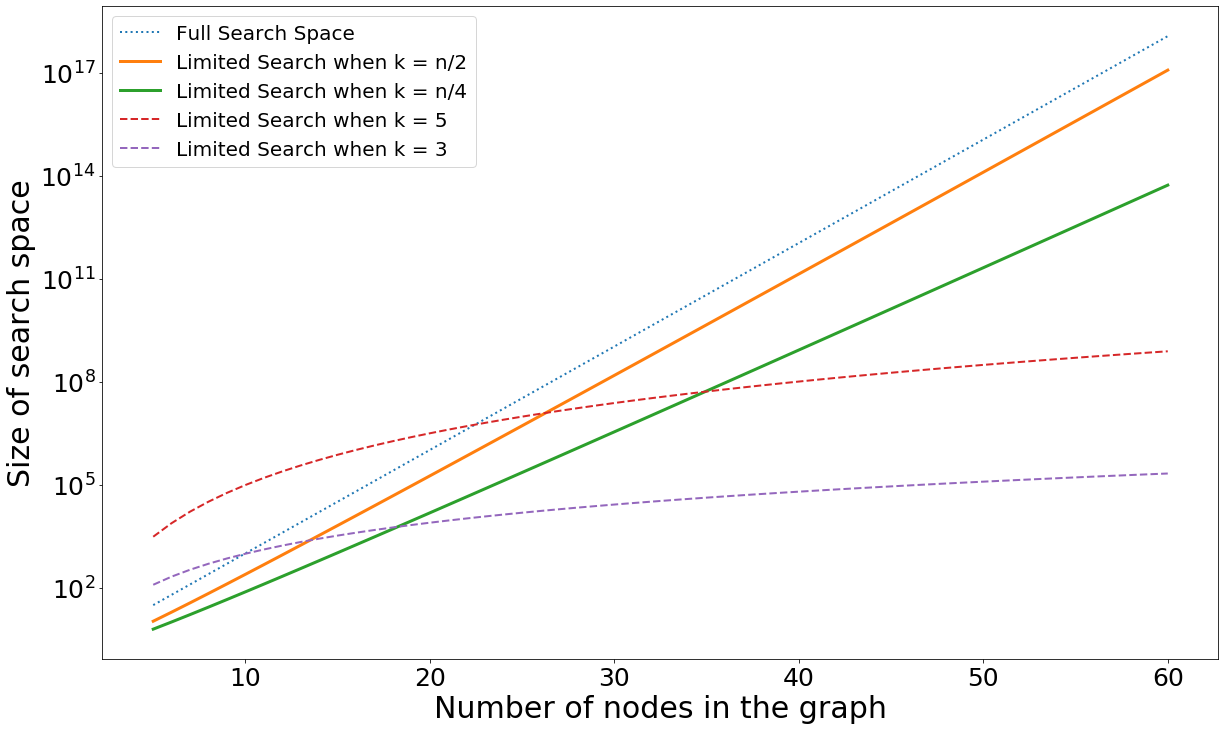}}
\caption{The difference in search space with respect to number of nodes in the graph. A limited search space (the two solid lines) is produced using state preparation (Dicke/ W states) in the cases $k=n/2$ and $k=n/4$.}
\label{ss}
\end{figure}

Fig.~\ref{ss} shows the change in the search space size for different clique sizes and approaches as the number of nodes grows. In the figure, the x-axis represents the number of nodes in the graph, and the y-axis represents the size of the search space. The worst-case search space for subsets of $n$ nodes is $2^n$ (upper dotted line). For fixed-size cliques, simple search methods are polynomial (lower dashed lines, $k=3$ and $k=5$). When the clique
size is a function of $n$, the search space is superpolynomial, and classical search becomes impractical.  Constrained-Hamming-weight quantum searches using Dicke states extend the range of problems that can be addressed using Grover’s algorithm (solid lines, $k = n/4$ and $k=n/2$).
%As can be seen, both increase exponentially, yet the limited search space increases with a slower pace than the full search space. The figure is generated for the worst case of limited search space, which occurs when $k=n/2$. In addition, the figure shows the effect of limiting the search space using state preparation in case of $k=3$ (triangle finding) for different size graphs.

\subsection{Checking-based Oracle}
To determine that a triangle exists, we need to confirm that 3 nodes are connected with 3 edges. This counting of nodes and edges is exactly what the oracle circuit should do.
In the checking-based oracle, each node in the graph is represented as a qubit, and the edges between them are expressed using one or more multiple-Toffoli $C^{\otimes n}X$ gates connecting specific qubits. After all edges have been counted, the results are checked. The sequence of $C^{\otimes n}X$ gates forms a simple adder that adds one every time an edge is encountered. In the case of a triangle, after the $C^{\otimes n}X$ gates, we need to check that we have precisely three edges ($11_{2}$). To check for $11_{2}$, we need two qubits that we will call {\tt edges\_counter}.\\

\begin{figure}[htbp]
\centerline{\includegraphics[scale=0.59]{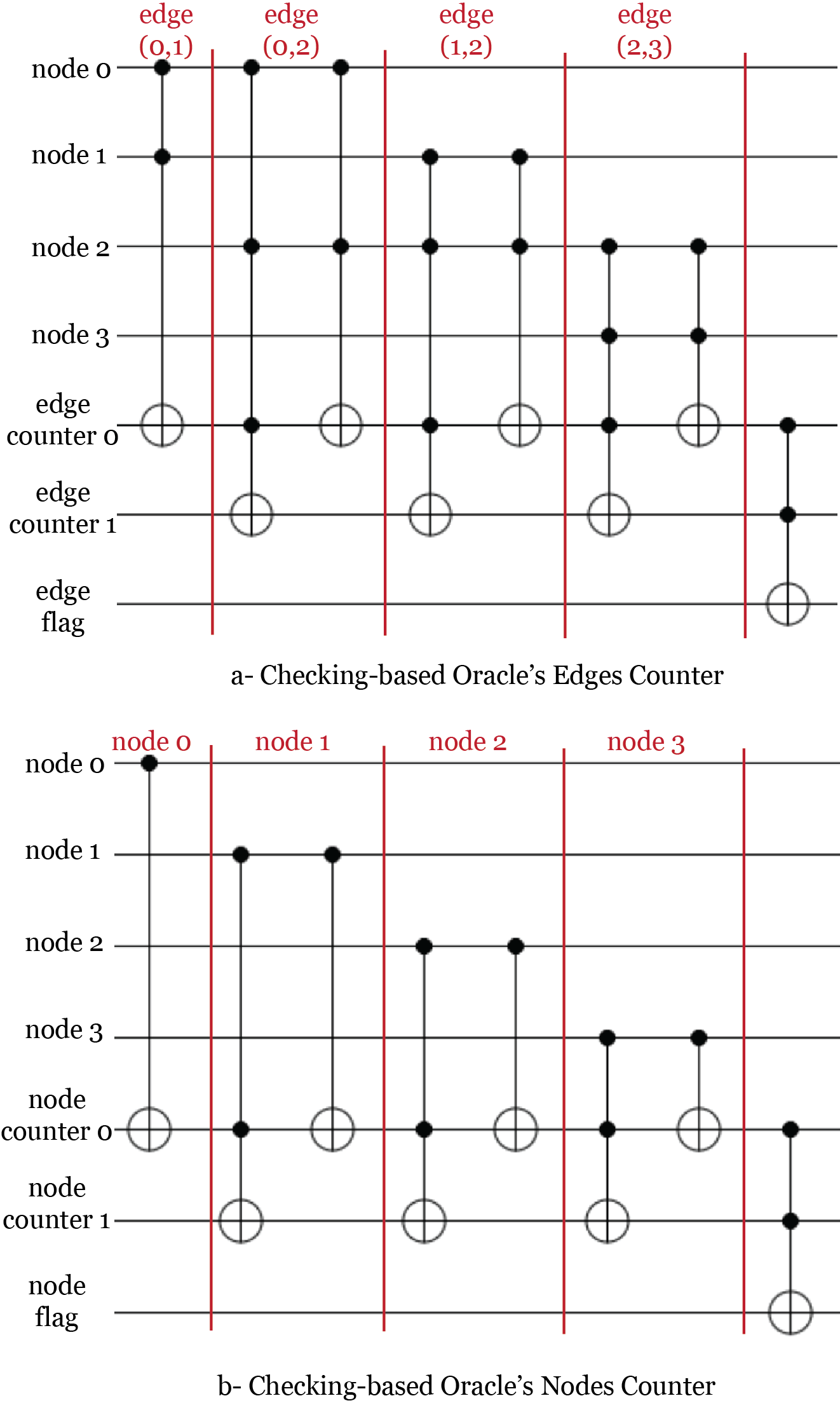}}
\caption{Checking-based Oracle for the graph in Fig.~\ref{exp_graph}.}
\label{orc1}
\end{figure}

\begin{figure}[htbp]
\centerline{\includegraphics[scale=0.5]{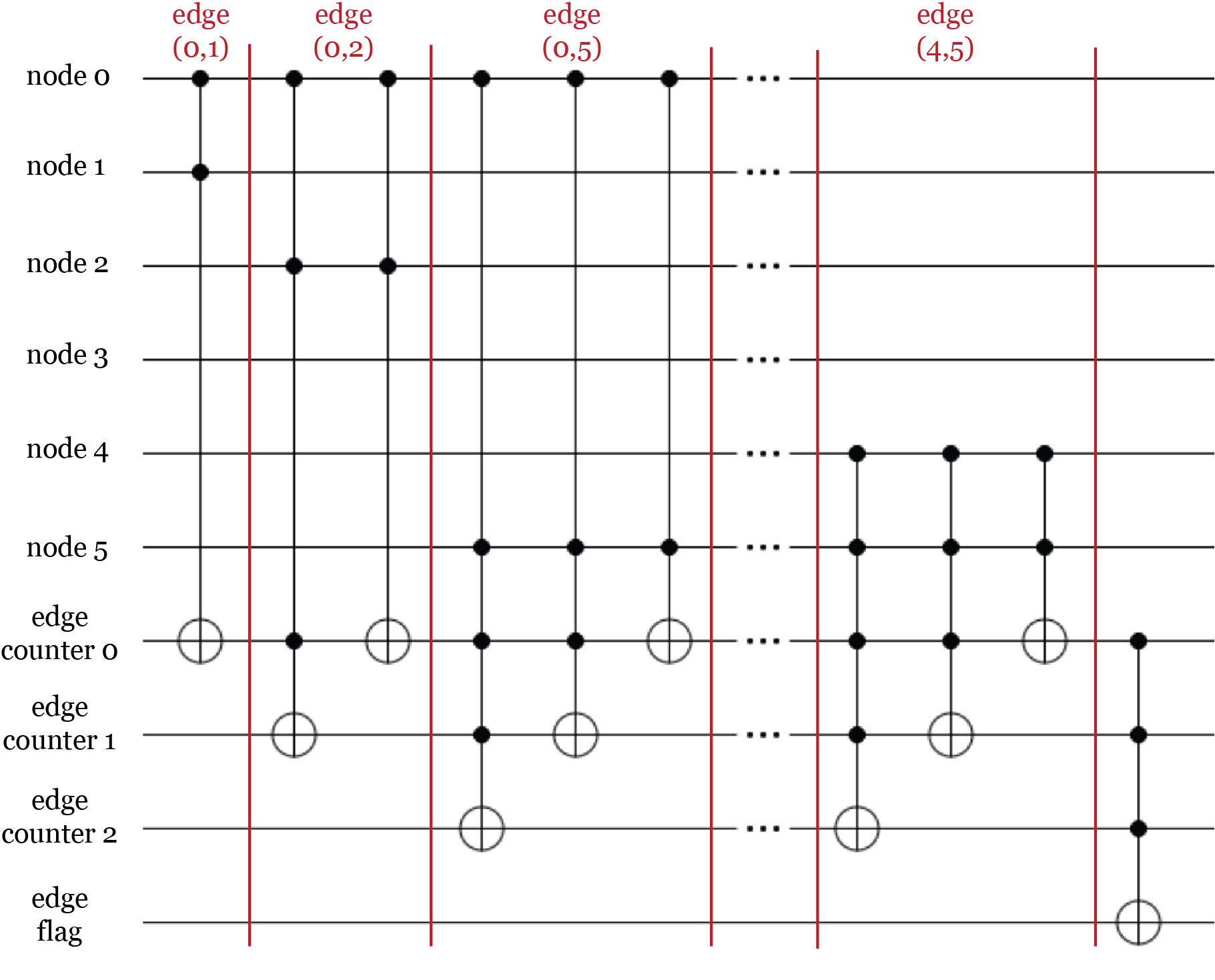}}
\caption{Checking-based Oracle for the graph in Fig.~\ref{graph}. The ten edges in the graph are expressed as ten groups of gates, giving the oracle cost $O(|E|)$.}
\label{orc-graph}
\end{figure}

In general, we need 
\(\lceil\log\binom{k}{2}\rceil\)
qubits to represent the {\tt edges\_counter}. For example, for 4-clique, the {\tt edges\_counter} will be a 3-qubit counter than can count up to 7 ($111_{2}$), and for a 5-clique which can count up to 15 ($1111_{2}$), the {\tt edges\_counter} will need four qubits and so on. For example, if we want to construct the oracle for the graph in Fig.~\ref{graph}, we will need six node qubits, a 3-qubit edge counter, and one qubit edge flag. The connection of the edges in the graph is then made, as shown in Fig.~\ref{orc-graph}.

Finally, to check if the {\tt edges\_counter} contains the correct value, another $C^{\otimes n}X$ gate needs to be applied, the result of which will be saved in another qubit,  {\tt edge\_flag} (Fig.~\ref{orc1}-a).
A similar circuit is then applied to count nodes; a $k$-clique should have $k$ nodes. The {\tt node\_counter} needs $\lceil \log k \rceil$ qubits with $C^{\otimes n}X$ between them. If the {\tt node\_counter} contains the correct number of nodes ($k$), the qubit {\tt node\_flag} will become 1. Fig.~\ref{orc1}-b shows the node counting section of the oracle. 
Finally, after checking for both edges and nodes, a \textsc{ccx} is applied to {\tt edge\_flag} and {\tt node\_flag} and stored in another qubit {\tt clique\_exists}. If we have the correct number of both the edges and the nodes, then a clique of size $k$ exists; otherwise, no clique exists.

\subsection{Incremental -based Oracle}
\begin{figure}[htbp]
\centerline{\includegraphics[scale=0.5]{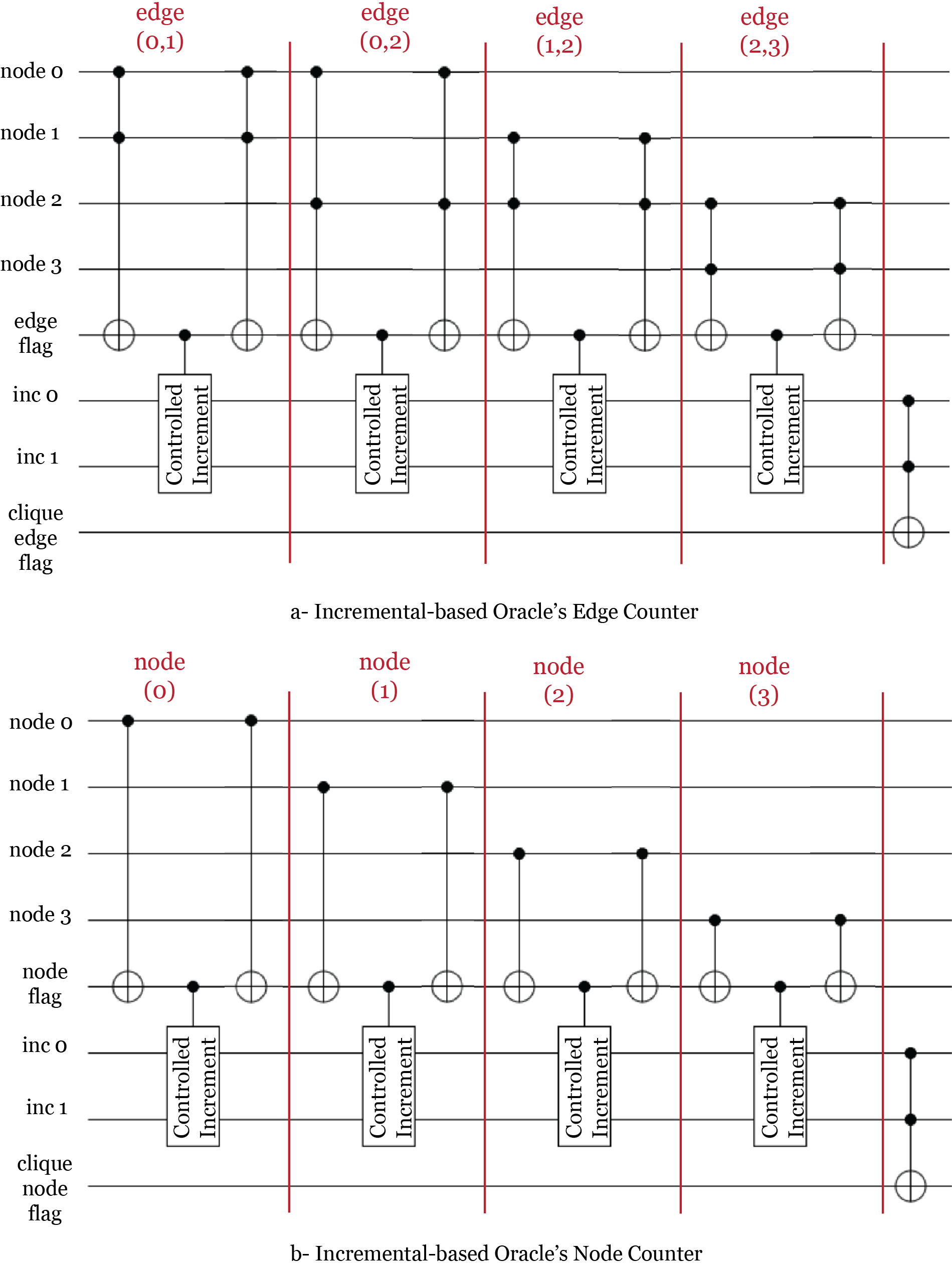}}
\caption{Incremental-based Oracle for the graph in Fig.~\ref{exp_graph}.}
\label{orc2}
\end{figure}

For incremental-based oracle, each node in the graph is represented with a qubit, and the edges are expressed using $C^{\otimes n}X$ gates. The difference between this and the checking-based oracle is in the {\tt edges\_counter} and {\tt clique\_flag}. In this implementation, the {\tt edges\_counter} is replaced with a one qubit {\tt edge\_flag}, the {\tt edge\_flag} becomes 1 if and only if an edge exists between two nodes. That flag is then used as a control qubit controlling an increment circuit that adds one every time it encounters an edge (Fig.~\ref{orc2}). In order for the {\tt edge\_flag} to function correctly, we need to uncompute it (reset to $|0\rangle$ state) after each increment.

\begin{figure}[htbp]
\centerline{\includegraphics[scale=0.33]{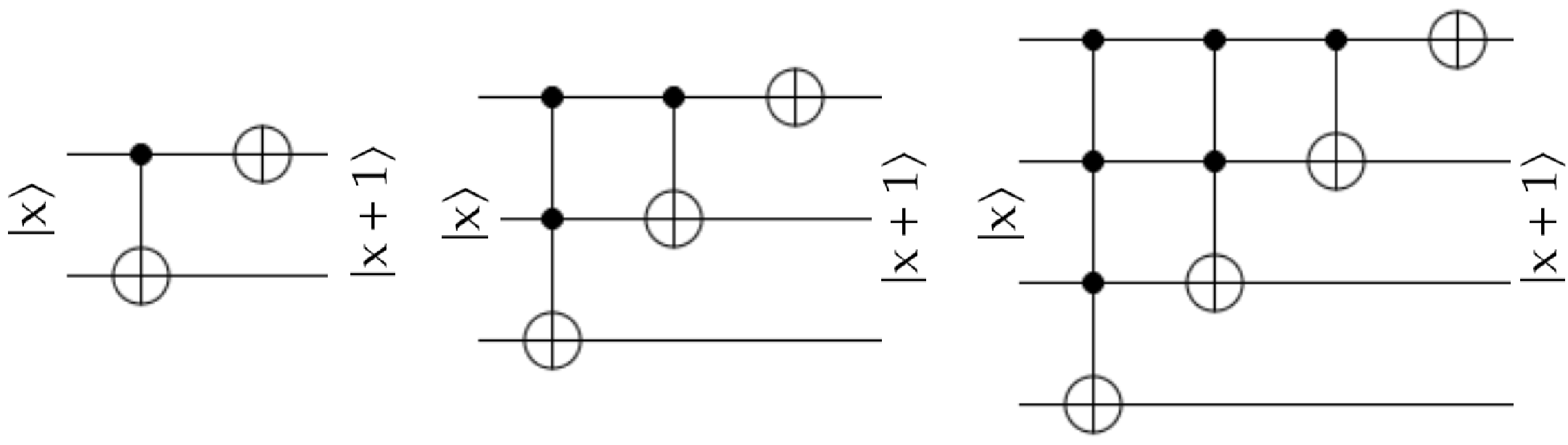}}
\caption{Different Size Increment Circuits. From right to left, 2-qubit increment, 3-qubit increment and 4-qubit increment circuits.}
\label{inc}
\end{figure}

The increment circuit size depends on the clique size; it will need \(\lceil\log\binom{k}{2}\rceil\) qubits. For example, when applying the oracle for a triangle ($k$ = 3), we will need a $2$-qubit increment circuit to count up to 3 or $11_{2}$. Fig~\ref{inc} shows different sizes of the increment circuit.
The circuit for finding the triangle in Fig.~\ref{exp_graph} needs two qubits for the increment circuit in addition to some ancillary qubits to implement the control functionality. The 2-qubit increment circuit, 3-qubit increment circuit, and 4-qubits increment circuits can be seen in Fig.~\ref{orc2}-a. After finishing all the edges in the graph, the qubit {\tt clique\_edge\_flag} will be $1$ only if the number of edges is correct \(\binom{k}{2}\).
When applying this oracle on the entire Hilbert Space, another circuit to count nodes needs to be added to the oracle~\ref{orc2}-b. The size of the controlled increment circuits in the case of counting nodes is $\lceil \log k \rceil$. The number of nodes will be stored in qubit {\tt clique\_node\_flag}.
Once both the edge counter and the node counter sections of the oracle are executed, the {\tt clique\_edge\_flag} and {\tt clique\_node\_flag} are used in a \textsc{ccx} to generate the {\tt clique\_flag} which will indicate if a clique of size $k$ exists in the graph or not.

\section{Analysis}
\label{res}

In order to test the efficiency of our implementation, we compared various combinations of the problem variables. To be consistent, the comparison is based on the smallest instance of the problem, i.e. the triangle finding problem. The combinations in the comparison are:
\begin{itemize}
\item Grover's algorithm with checking-based oracle over the entire Hilbert space.
\item Grover's algorithm with checking-based oracle over limited search space using W-state preparation (W state followed by $n$ NOT gates).
\item Grover's algorithm with checking-based over limited search space using Dicke state preparation.
\item Grover's algorithm with incremental-based oracle over the entire Hilbert space.
\item Grover's algorithm with incremental-based oracle over limited search space using W state preparation (W state followed by $n$ NOT gates).
\item Grover's algorithm with incremental-based over limited search space using Dicke state preparation.
\end{itemize}
We will address the analysis from two perspectives, complexity and practicality, comparing the type of gates and depth of the resultant circuit. In addition, we will also discuss how different state preparation affects the amplitude of the correct answer, using both the ideal-case simulation and gate-error simulation.

\subsection{Gate Count Analysis}
In quantum circuits, the more gates that involve multiple qubits, the more unreliable and difficult it will be to get the circuit to work on actual quantum hardware. 
First, we will discuss the different circuit sizes for different implementations of the oracle and various state preparations. Again, as a base case, we will compare the different approaches in the case of finding a 3-clique (triangle) in a 4-node graph. Table~\ref{tab:size-depth} shows the different operation counts from both checking-based and incremental-based oracle for the optimal oracle iteration count.

\begin{table}[]
\centering
\caption{Circuit size, depth (length of critical path) and number of qubits needed for all approaches of Checking-based oracle and Incremental-based oracle for the optimal number of iterations for the triangle finding problem in Fig.~\ref{exp_graph}}
\label{tab:size-depth}
\resizebox{7.5cm}{!}{%
\begin{tabular}{|
>{\columncolor[HTML]{FFFC9E}}c |c|c|c|}
\hline
\multicolumn{4}{|c|}{\cellcolor[HTML]{C0C0C0}Checking-based Oracle}                                                                                    \\ \hline
\cellcolor[HTML]{333333} & \cellcolor[HTML]{FFFC9E}Full Search Space & \cellcolor[HTML]{FFFC9E}W state Prep & \cellcolor[HTML]{FFFC9E}Dicke state Prep \\ \hline
Size                     & 214                                       & 97                                   & 259                                      \\ \hline
Depth                    & 165                                       & 79                                   & 237                                      \\ \hline
\# of Qubits             & 13                                        & 9                                    & 9                                        \\ \hline
\multicolumn{4}{|c|}{\cellcolor[HTML]{C0C0C0}Incremental-based Oracle}                                                                                 \\ \hline
\cellcolor[HTML]{333333} & \cellcolor[HTML]{FFFC9E}Full Search Space & \cellcolor[HTML]{FFFC9E}W state Prep & \cellcolor[HTML]{FFFC9E}Dicke state Prep \\ \hline
Size                     & 1000                                      & 215                                  & 281                                      \\ \hline
Depth                    & 837                                       & 131                                  & 200                                      \\ \hline
\# of Qubits             & 15                                        & 10                                   & 10                                       \\ \hline
\end{tabular}%
}
\end{table}

To understand better the numbers in Table~\ref{tab:size-depth}, we need to consider how many times the oracle is repeated. Since Grover's Algorithm is periodic, the optimal number of repetitions of the oracle and diffusion is calculated based on the number of input qubits (number of nodes in the graph in our case) and the number of solutions we want the algorithm to find as shown in Eq.~\ref{iter}. 
If we are using the entire search space, then $N = 2^n$ while $m = 1$, and so the optimal number of iterations here will be three iterations. However, if we are using Dicke/ W states to limit our search space, $N$= \(\binom{n}{k}\), $m = 1$ giving an optimal iteration number of one. 
Although the number of iterations is smaller with state preparation (Dicke/ W state), the circuit may increase in size, based on the state preparation approach followed.%bigger than searching the entire search space. 
While it is relatively easy to prepare initial states in full superposition (only $n$ H gates), preparing initial state using Dicke/ W states is a costly operation, with Dicke state being the most expensive in gate count. Detailed layout of the gates used in every state preparation is found in Table~\ref{tab:stat-size}.

\begin{table}[htbp]
\centering
\caption{Gate type and count for each state preparation approach}
\label{tab:stat-size}
\resizebox{\columnwidth}{!}{%
\begin{tabular}{|c|c|c|}
\hline
\rowcolor[HTML]{C0C0C0} 
{\color[HTML]{333333} State Preparation Method} & {\color[HTML]{333333} Gate Count} & {\color[HTML]{333333} Gate Type} \\ \hline
Full search space & 4 & (Hadamard, 4) \\ \hline
W State & 17 & (U3, 6), (CNOT, 6), (NOT, 5) \\ \hline
Dicke State & 39 & (CNOT, 18), (U3, 12), (CCNOT, 6), (NOT, 3) \\ \hline
\end{tabular}%
}

\end{table}

Circuit size by itself is crucial, but it is more important to check the full list of gates used. More particularly, NOT, CNOT, CCNOT, C$^{\otimes N}$NOT gate counts play an essential factor in whether the circuit can be applied to a real hardware device or not. Table~\ref{tab:not-count} lists the number of NOT, CNOT, $C^{\otimes N}NOT$ gates in every approach proposed for the optimal number of iterations for each.

Another factor affecting whether a circuit is implementable or not is the circuit depth or length of the critical path of the circuit. Unfortunately, the circuit depth is highly dependent on the hardware layout of qubits and the connections between them. On the bright side, many works have focused on optimizing and generalizing circuits depth and size for any hardware qubit layout~\cite{childs2019circuit}~\cite{andres2019automated}~\cite{Siraichi:2018:QA:3179541.3168822}~\cite{nishio20:jetc-pre}~\cite{Nash_2020}.

\begin{table}[]
\centering
\caption{The number of NOT, CNOT and CCNOT gates in Checking-based and Incremental-based approaches for the triangle finding problem in Fig.~\ref{exp_graph}}
\label{tab:not-count}
\resizebox{7.5cm}{!}{%
\begin{tabular}{|
>{\columncolor[HTML]{FFFC9E}}c |c|c|c|}
\hline
\multicolumn{4}{|c|}{\cellcolor[HTML]{C0C0C0}Checking-based Oracle}                                                                                    \\ \hline
\cellcolor[HTML]{333333} & \cellcolor[HTML]{FFFC9E}Full Search Space & \cellcolor[HTML]{FFFC9E}W state Prep & \cellcolor[HTML]{FFFC9E}Dicke state Prep \\ \hline
NOT                      & 25                                        & 24                                   & 18                                       \\ \hline
CNOT                     & 24                                        & 18                                   & 90                                       \\ \hline
CCNOT                    & 123                                       & 31                                   & 37                                       \\ \hline
\multicolumn{4}{|c|}{\cellcolor[HTML]{C0C0C0}Incremental-based Oracle}                                                                                 \\ \hline
\cellcolor[HTML]{333333} & \cellcolor[HTML]{FFFC9E}Full Search Space & \cellcolor[HTML]{FFFC9E}W state Prep & \cellcolor[HTML]{FFFC9E}Dicke state Prep \\ \hline
NOT                      & 25                                        & 24                                   & 18                                       \\ \hline
CNOT                     & 312                                       & 42                                   & 78                                       \\ \hline
CCNOT                    & 99                                        & 17                                   & 35                                       \\ \hline
\end{tabular}%
}
\end{table}

\subsection{Simulation Results Analysis}
This subsection discusses how the change in state preparation affects the amplitude of the correct answer (probability of success). To observe this change, we will simulate the circuit twice, once using the ideal-case simulator (QASM Simulator) and another simulation with added gate error.
The Qiskit Aer module provides the pure-state QASM simulator. Aer is a high-performance Qiskit simulation framework for quantum circuits. It offers various backends to meet different simulation ends. QASM simulates any given circuit assuming ideal qubits and gates with no errors. The results of using the QASM simulation are not realistic for current hardware, and represent the goal of future advancements in quantum computers. However, for now, ideal simulators are used. For more realistic results, Aer also provides a way to add noise to the gates while assuming perfect qubits. In real life, both qubits and gates are faulty and noisy, but adding gate noise produces more realistic simulation results.\\

\subsubsection{Thermal-relaxation Error}
\label{thermal}

There are several types of errors that can be applied to the QASM simulator; namely, Qiskit Aer offers ten standard error models, including Depolarization Error, Reset Error, and Thermal Error with an option to create user-customized error models~\cite{aleksandrowicz2019qiskit}.
 In addition, the user can choose whether to apply the error to all qubits or a specific set of qubits. In our gate-error simulation, we decided on a realistic thermal-error model (thermal relaxation) and applied it to all qubits in the algorithm. \\
Thermal relaxation needs two main parameters defined, $T_1$ and $T_2$, together called decoherence times. $T_1$ is known as the relaxation time constant; it is defined as the time needed for the system to go from state $|1\rangle$ to $|0\rangle$ with probability $\frac{1}{e}$. $T_2$ is defined as the dephasing time constant, expressing how long the phase of certain qubit stays intact, in other words, it is the time from state $|-\rangle$ to state $|+\rangle$ with probability $\frac{1}{e}$~\cite{hu2002decoherence}. %Additionally, $T_1$ represents the decay time for any state, for example.
%the probability of state $|\psi\rangle$ remaining at its state for some time t is given by Eq.~\ref{decay}. 
%On the other hand, $T_2$ represents the autocorrelation time of which state $|-\rangle$ and state $|+\rangle$ become uncorrelated, meaning having a mixed state from which the initial state cannot be determined.\\

%\begin{equation}
%P( |\psi\rangle ) =e^{−\frac{t}{T_1}}
%\label{decay}
%\end{equation}

%%
%we should point out that the error rates are determined by gate execution times and the qubit $T_1$ and $T_2$ values. The values chosen for the gate execution times are averages based on actual devices as follows, U2 gates take 50 nanoseconds, U3 gates take 100 nanoseconds, CNOT gates take 300 nanoseconds, and finally, the readout will take 1000 nanoseconds\footnote{U2, and U3 are basic single-qubit unitary gates presented by Qiskit~\cite{elementaryqiskit}}.
%The effect of applying these values can be seen in Fig.~\ref{err}.

\begin{table}[]
\centering
\caption{Average values of $T_1$ and $T_2$ in microseconds for six different IBMQ Devices}
\label{tab:dev_time}
\resizebox{6cm}{!}{%
\begin{tabular}{|
>{\columncolor[HTML]{FFFC9E}}c |c|c|}
\hline
\cellcolor[HTML]{C0C0C0}\begin{tabular}[c]{@{}c@{}}Device \\ Name\end{tabular} &
  \cellcolor[HTML]{C0C0C0}\begin{tabular}[c]{@{}c@{}}$T_1$ \\ (in $\mu$s)\end{tabular} &
  \cellcolor[HTML]{C0C0C0}\begin{tabular}[c]{@{}c@{}}$T_2$  \\ (in $\mu$s)\end{tabular} \\ \hline
ibmq\_melbourne    & 55 & 59 \\ \hline
imbq\_poughkeepsie & 64 & 65 \\ \hline
ibmq\_singapore    & 83 & 89 \\ \hline
ibmq\_paris        & 76 & 67 \\ \hline
ibmq\_cambridge    & 81 & 39 \\ \hline
ibmq\_rochester    & 55 & 59 \\ \hline
\end{tabular}%
}
\end{table}

\begin{figure}
\centerline{\includegraphics[scale=0.5]{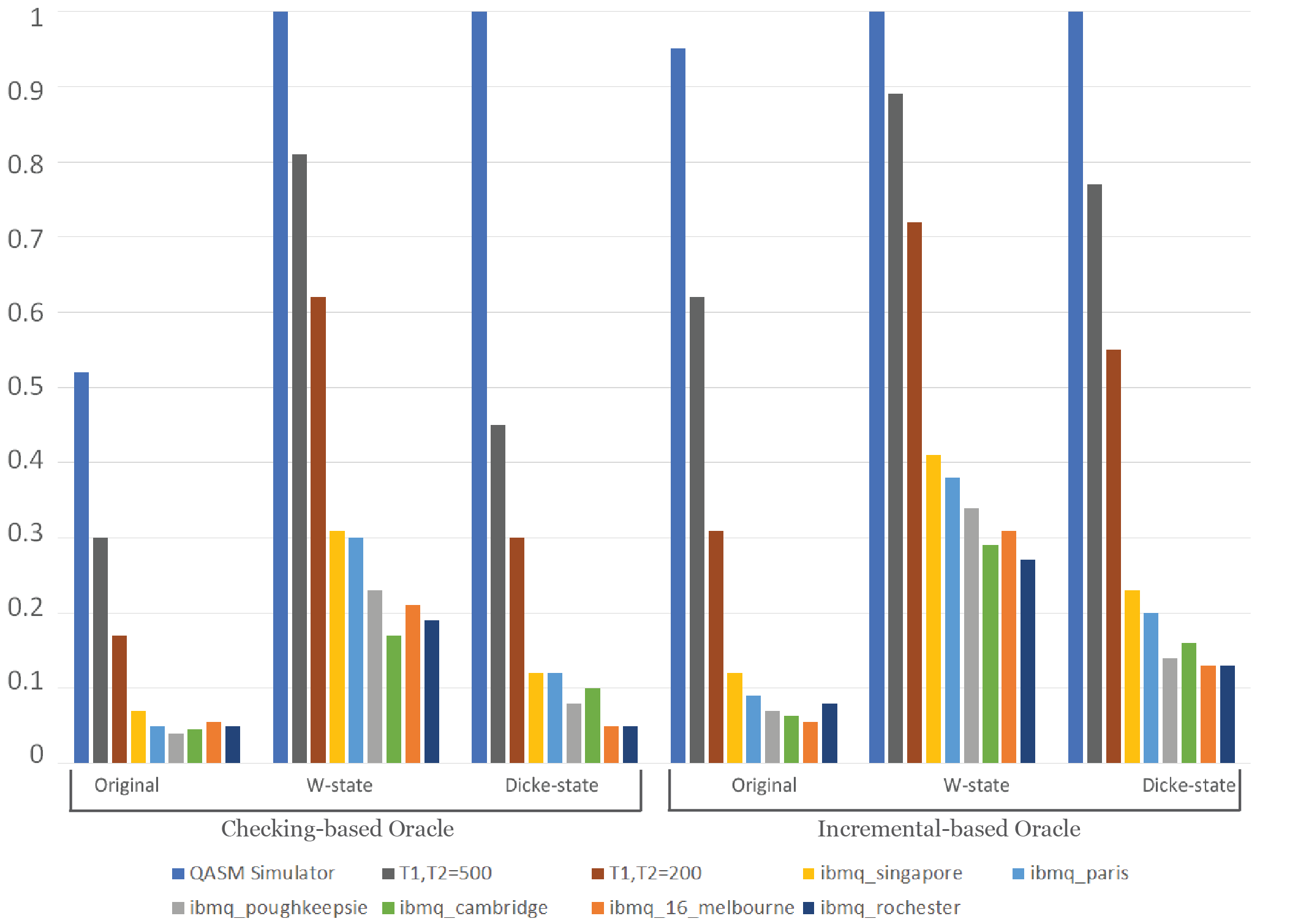}}
\caption{The amplitude damping effect of memory decoherence, assuming perfect gates. The bars are the probability of finding the correct answer after simulating a perfect machine (leftmost bar in each group), $T_1 = T_2$ = 500, 200 (next two bars) as well as $T_1$ and $T_2$ based on the six different IBMQ devices in table~\ref{tab:dev_time} (last six bars). The figure is sorted based on average error rate from lowest to highest.}
\label{dev_sims}
\end{figure}

In order to understand better the relation between $T_1$, $T_2$, and how they affect the amplitude of the correct answer, we applied our two proposed oracle structures (for the graph in Fig.~\ref{exp_graph}) to six different IBMQ devices with different $T_1$, $T_2$. Since the value of $T_1$, $T_2$ depends on the specific qubits; we took the average $T_1$, $T_2$ of the devices when we applied our different circuits. Table~\ref{tab:dev_time} shows the average values of $T_1$, $T_2$, and the names of the six devices used.
We should point out that the error rates are determined by gate execution times and the qubit $T_1$ and $T_2$ values. The values chosen for the gate execution times are averages based on actual devices as follows, U2 gates take 50 nanoseconds, U3 gates take 100 nanoseconds, CNOT gates take 300 nanoseconds, and finally, the readout will take 1000 nanoseconds\footnote{U2, and U3 are basic single-qubit unitary gates presented by Qiskit~\cite{elementaryqiskit}}.
Fig.~\ref{dev_sims} shows the results of all proposed approaches on each of the six devices. Various observations can be made by looking at the bar chart. Mainly, it can be seen that the W-state preparation approach retains the correct answer better than other methods, followed by the incremental-based Dicke state preparation approach. It can also be seen that the {\tt ibmq\_singapore} device has the lowest error among this set of devices, followed by {\tt ibmq\_paris}, which is due to these devices having the highest $T_1$, $T_2$ among the devices used. In addition, we added another simulation where $T_1$, $T_2$ = 200 $\mu$s, and 500 $\mu$s. As can be seen from the figure, these simulations have the least error rate among all simulations performed. Hence, increasing $T_1$, $T_2$ by 60\% reduced the error rate and the damping in the amplitude of the correct answer by nearly 42\%.

\subsubsection{Device-specific Error}
\label{device}
The above case incorporates only memory errors; gates are assumed perfect. Hence, to provide a more realistic effect of noise models in NISQ devices, we applied the device-specific noise models to three of our implementations. The three implementations we chose to apply device-specific models are Checking-based Oracle with W-state Preparation, Incremental-based Oracle with W-state Preparation, Incremental-based Oracle with Dicke state Preparation.
We chose these three approaches because they have the highest error tolerance among the six strategies. All three implementations have nine qubits circuits and an ideal (QASM simulator) amplitude of 1. 

\begin{figure}
\centerline{\includegraphics[scale=0.5]{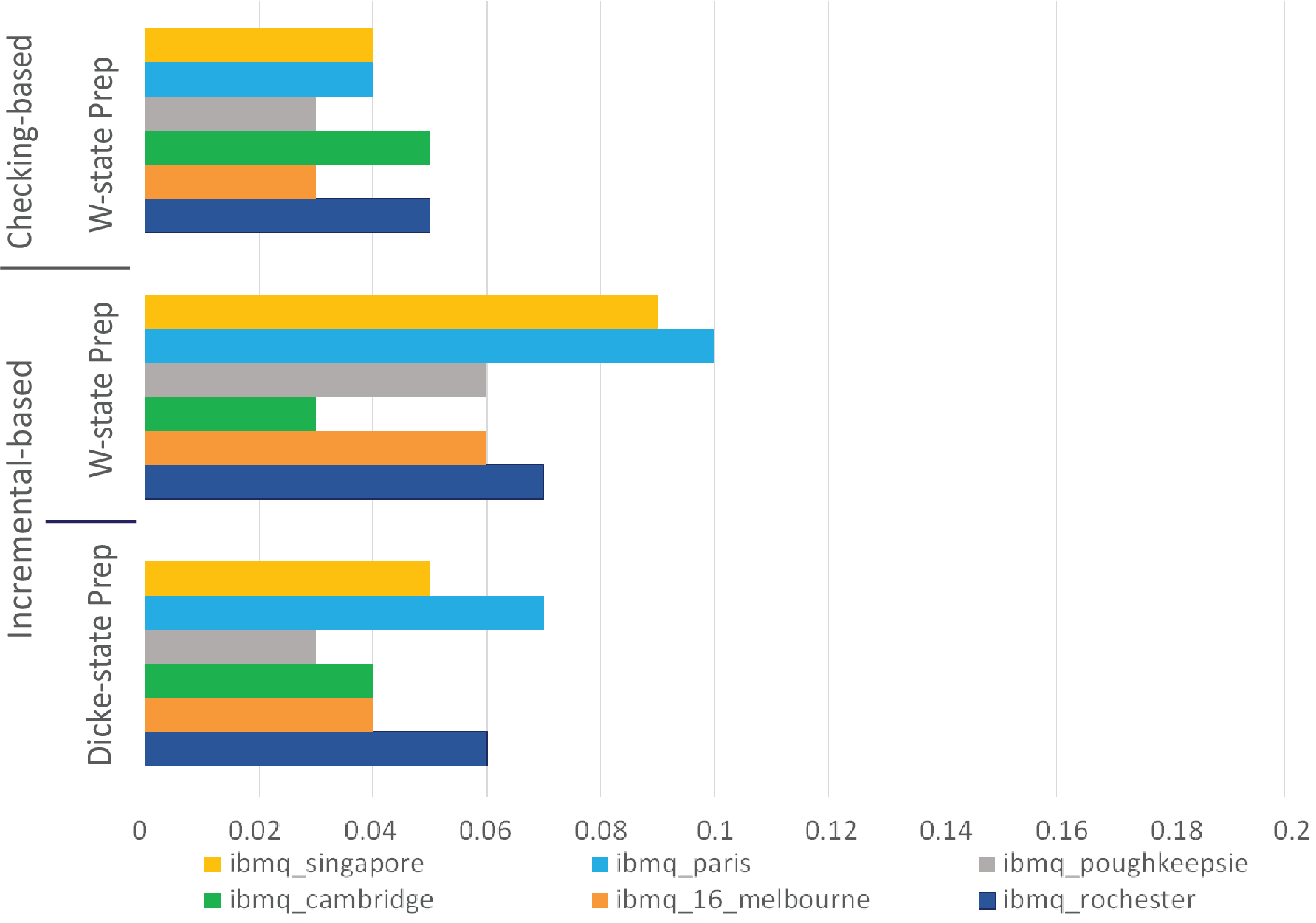}}
\caption{Probability of finding the correct answer using the Checking-based Oracle with W-state Preparation, Incremental-based Oracle with W-state Preparation, and Incremental-based Oracle with Dicke state Preparation. Data taken on the six different IBMQ devices in table~\ref{tab:dev_time}.}
\label{dev-sep-err}
\end{figure}

%Considering Fig.~\ref{dev-sep-err}, we can observe that when applying the full noise model of any of the IBMQ devices, the error rate increases sharply. Even the implementations with high error tolerance for changes in $T_1$, $T_2$, show a significant drop in the amplitude of the correct answer, with error rate, range of 93\% to 96\%. We can also see that, although {\tt ibmq\_singapore} and {\tt ibmq\_paris} showed the best performance when manipulating  $T_1$, $T_2$, when applying full noise model, these two devices still showed better performance than the rest of the machines.

Considering Fig.~\ref{dev-sep-err}, we can observe that when executing the Checking-based Oracle with W-state Preparation, Incremental-based Oracle with W-state Preparation, and Incremental-based Oracle with Dicke state Preparation on real IBMQ devices, the error rate increases sharply. Even the implementations with high error tolerance for changes in $T_1$, $T_2$, show a significant drop in the amplitude of the correct answer, with error rate ranging from 93\% to 96\%. We can also see that {\tt ibmq\_singapore} and {\tt ibmq\_paris} maintained the best performance among the six devices used.
We compared running the approaches on the real devices to simulating the devices' error models on the QASM simulator and found that the results are incredibly close, with negligible differences. That gave us confidence that using the error models of the devices provides a valid representation of the performance of the actual devices.

\subsection{Time Complexity Analysis}
\label{time analysis}
%The $k$-clique problem is considered an NP-complete problem with time complexity $O(n^k)$. If $k$ is fixed and small, the complexity becomes polynomial, and if $k$ is large, then the time complexity is expected to be polynomial; for example, in the case of the triangle finding $k=3$, the complexity becomes $O(n^3)$.
For our proposed approaches, we can split the time complexity analysis into four main parts: analyzing the number of iterations in Grover's algorithm, 
%the algorithm steps in order Grover's overall complexity, 
the initial state preparation (in case of limited Hilbert space search) complexity, the different oracles and diffusion operators complexities, and finally analyzing the total complexity of the algorithm.
%the complexity of the diffusion operator.

\subsubsection{Number of Iterations in Grover's Algorithm}
The oracle and the diffusion operator are repeated $\lfloor\frac{\pi}{4}\sqrt{\frac{N}{m}}\rfloor=O(\sqrt{\frac{N}{m}})$ times, which depends on the size of the search space and the expected number of answers. Assuming the simplest case, where $m=1$, such as the case in Fig.~\ref{exp_graph}, the complexity then becomes $O(\sqrt{N})$. Notice that applies to the case when the entire Hilbert space is used, however, if we limit the search space using initial state preparation, the number of iterations then depends also on the size of clique $k$ and becomes \(O(\sqrt{\binom{n}{k}})\).

\subsubsection{State Preparation Complexity}
In section~\ref{imp}, we proposed the usage of two different state preparation techniques to limit the search space. Using either W-state preparation in case $k=n-1$ or using Dicke state preparation otherwise. As mentioned in section~\ref{bg}, we followed the algorithm in~\cite{cruz2019efficient} to prepare the nodes qubits in a W-state superposition; the algorithm produces a circuit with depth $O(\log n)$ and complexity of $O(n)$. Here $n$ represents the number of qubits involved in the W-state preparation, which is, in our case, the number of nodes $|V|$. Hence the cost of preparing W-states becomes $O(|V|)$. 
On the other hand, when using the Dicke state preparation proposed in~\cite{bartschi2019deterministic}, we get a circuit with complexity $O(kn)$ and depth $O(n)$, where $k$ is the clique size, and $n$ is the number of qubits. Therefore, the cost of preparing the Dicke state becomes $O(k|V|)$.\\

\subsubsection{Oracle and Diffusion Operator Complexities}
First, we will discuss the complexity of the diffusion operator. As can be seen in Fig.~\ref{diff}, the diffusion operator consists of the adjoint of the state preparation, a $C^{\otimes n}Z$ gate, and a state preparation, respectively. Hence, we can generalize the complexity of the diffusion operator as $O(state\_prep)+O(C^{\otimes n}Z)+O(state\_prep)$. The cost of the state preparation depends on which approach is used; hence, it will be $O(|V|)$ in case of W-state preparation or $O(k|V|)$ in case of Dicke state preparation, as can be seen in Table~\ref{tab:time_comp}. However, the complexity of the $C^{\otimes n}Z$ gate depends on the number of nodes $|V|$, therefore the complexity of the gate will be $O(|V|)$. Consequently, the total complexity of the diffusion operator will become $O(state\_prep)+O(|V|)$.\\
The complexity of the oracle, however, depends on whether an initial state preparation is used. Regardless of the oracle implementation (checking-based or incremental-based), the primary function of the oracle counts the number of edges and nodes needed to compose a clique of size $k$, in addition to $k$ itself. So, the complexity of the oracle for the entire Hilbert space is $O(k+|V|+|E|)$. If we use state preparation, we are eliminating the need to count nodes; that is because we only allow states with the specific $k$ nodes activated at any time to be included in the search space, thus eliminating the need to count the nodes in the clique. Hence, the complexity of the oracle when using initial state preparation to limit the search space is $O(k+|E|)$.

\begin{table}[]
\centering
\caption{Different for the different steps of the algorithm with and without the initial state preparation}
\label{tab:time_comp}
\resizebox{6.5cm}{!}{%
\begin{tabular}{|
>{\columncolor[HTML]{FFFC9E}}c |c|c|}
\hline
\cellcolor[HTML]{C0C0C0}\begin{tabular}[c]{@{}c@{}}Algorithmic \\  Step\end{tabular} &
  \cellcolor[HTML]{C0C0C0}\begin{tabular}[c]{@{}c@{}}Circuit\\ Depth\end{tabular} &
  \cellcolor[HTML]{C0C0C0}\begin{tabular}[c]{@{}c@{}}Total\\ Complexity\end{tabular} \\ \hline
\begin{tabular}[c]{@{}c@{}}W state  \\ preparation\end{tabular}       & $O(\log |V|)$ & $O(|V|)$ \\ \hline
\begin{tabular}[c]{@{}c@{}}Dicke state  \\ Preparation\end{tabular}   & $O(|V|)$                    &  $O(k|V|)$\\ \hline
\begin{tabular}[c]{@{}c@{}}Diffusion  \\ Operator\end{tabular}        & $O(state\_prep)+ O(k|V|)$   &  $O(state\_prep)$\\ \hline
\begin{tabular}[c]{@{}c@{}}Oracle without \\  state prep\end{tabular} & $O(k+|V|)$               &$O(k+|E|)$  \\ \hline
\begin{tabular}[c]{@{}c@{}}Oracle with \\  state prep\end{tabular}    & $O(k+|E|)$                   & $O(k+|E|)$ \\ \hline
\end{tabular}%
}
\end{table}

%\begin{tabular}[c]{@{}c@{}}Grover's \\  iteration\end{tabular} & $O(\sqrt{\frac{\binom{n}{k}}{m}})$  &  $O(\sqrt{\frac{\binom{n}{k}}{m}})$ \\ \hline

\subsubsection{Algorithm Total Complexity}
The total complexity of Grover's algorithm can be expressed as the number of iterations times the cost of one iteration. The number of iterations, as discussed in previous subsections can be presented as $O(\sqrt{\frac{\binom{n}{k}}{m}})$. Each iteration's cost can be divided into two parts, the oracle's cost, and the diffusion operator's cost. Hence the total complexity becomes $O(\sqrt{\frac{\binom{n}{k}}{m}}) \times (O(\textrm{oracle})+O(\textrm{diffusion operator}))$. This complexity, however, assumes the initial state preparation of states in the entire Hilbert space. That would not be correct if we used W-state or Dicke-state as initial state preparation. In that case, the complexity becomes $O(\textrm{state\_prep})+O(\sqrt{\frac{\binom{n}{k}}{m}}) \times (O(\textrm{oracle})+O(\textrm{diffusion operator}))$.
\section{Discussion}
\label{dis}

To estimate the time when our proposed schemes of Grover's algorithm to solve the clique finding problem can be implemented on a real device with minimal error, we need to address two factors, the quantum volume, and the device performance.

\subsection{Quantum Volume}
\begin{figure}[htbp]
\centerline{\includegraphics[scale=0.35]{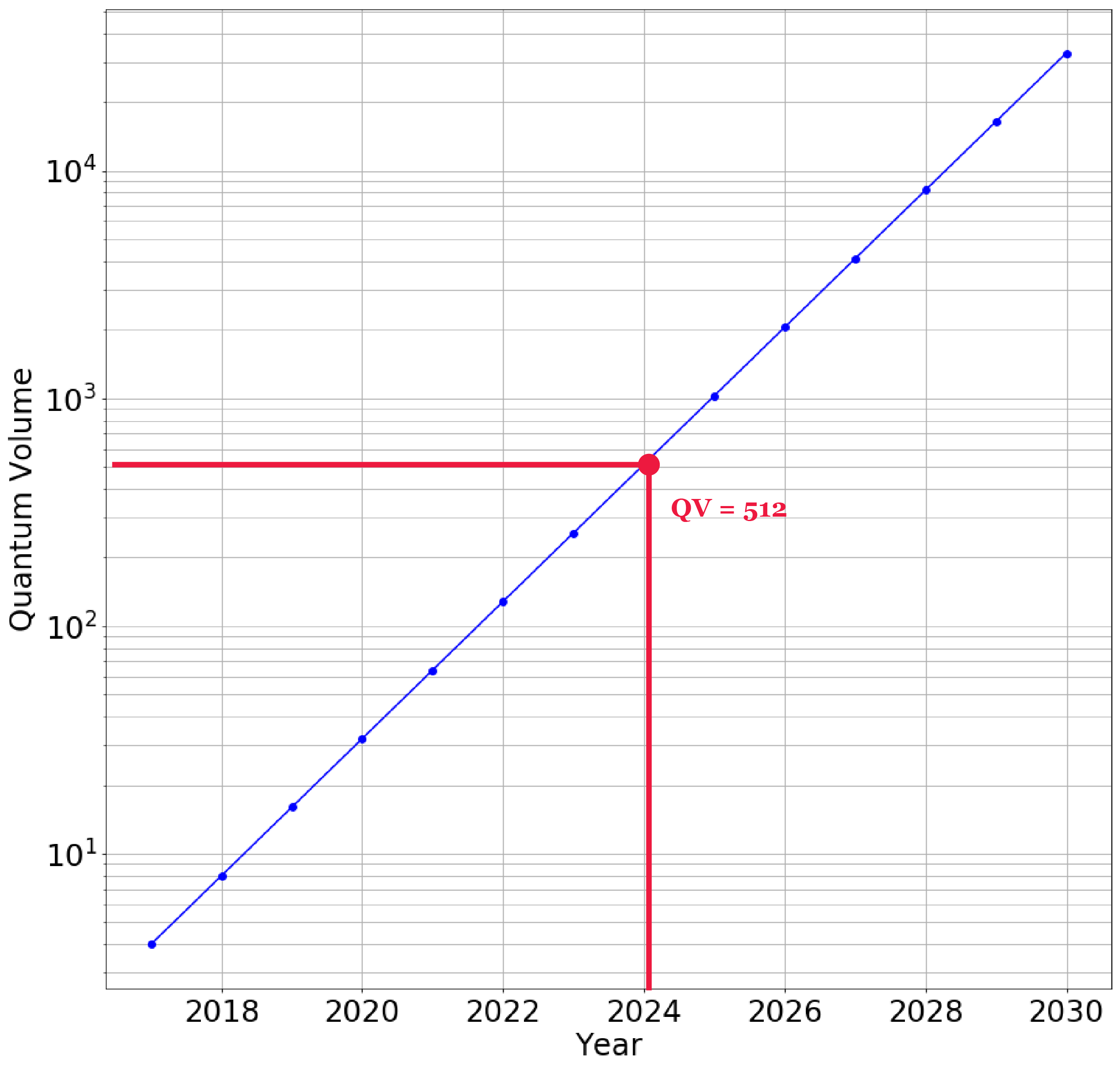}}
\caption{IBM's Quantum Volume Growth Chart with a highlight on 2024, when QV 512 is hoped to be achieved.}
\label{qv}
\end{figure}

IBM has proposed a single number indicator to describe the quantum processing capabilities of any NISQ device. IBM not only introduced the concept of Quantum Volume (QV)~\cite{cross2019validating}; they also laid out a prediction for the future of their quantum devices, Fig.~\ref{qv}. Their proposed roadmap for the advancement of quantum processor power aims to double the performance every year in order to achieve Quantum Advantage in the near future~\cite{gil2019future}. IBM devices reached QV of 4 in 2017, the same year they released the Qiskit library to the public. They followed that by entering QV of 8 in 2018, 16 with Q System One, and earlier this year (2020), they reached QV of 32.
This doubling in processing power is the quantum equivalent to Moore's Law, the famous observation by Gordon Moore about the exponential progress of classical computer chips.

QV represents the ability to run a circuit on an IBM quantum device with at least 2/3 probability of measuring an answer that passes some statistical test~\cite{cross2019validating}.

In order to run the smallest instance of our proposed scheme, which is the checking-based oracle with W-state preparation, on a real device with a reasonable probability of success, we need QV of at least 512. Since the size depth of that precise circuit at {\tt opt\_iter} are 97 and 79 respectively, and QV is calculated as described in Eq.~\ref{qv_qqn} introduced in~\cite{cross2019validating}, where $d$ represents the depth of the circuit and n is the number of qubits. 

\begin{equation}
Quantum Volume (QV) \approx 2^{min(d,n)}
\label{qv_qqn}
\end{equation}

QV of $2^9$ is needed to execute that circuit correctly. Hence, based on that and the growth chart proposed by IBM for future growth, it is expected to reach QV of 512 in early 2024, as highlighted in Fig.~\ref{qv}.

\subsection{Device Performance}
Even among machines with similar QV, their performance depends on more than just the number of qubits in the machine or the depth of the circuit that can be implemented on it. It also depends on the device noise model, as discussed in~\ref{device} and the coupling map (the connectivity between the qubits). There are various types of error, including the thermal relaxation error discussed in~\ref{thermal}, gate error, Pauli error, readout error, and measurement error. Each machine has a different amount of these errors per qubit, which makes it challenging to estimate the ability to implement any algorithm on a real device based solely on its QV.
We analyzed the performance of our top three error-resistant approaches (Checking-based Oracle with W-state Preparation, Incremental-based Oracle with W-state Preparation, Incremental-based Oracle with Dicke state Preparation) on the two machines with the overall best performance {\tt ibmq\_singapore} and {\tt ibmq\_paris}.
We obtained the error\_model of both these devices and modified it in three ways in order to understand which factor affects the overall error most. We changed the thermal relaxation error by modifying $T_1$, $T_2$ while keeping all other errors untouched, we then did the same but with gate error, and finally, we edited both the thermal relaxation error and the gate error together.

\begin{figure*}[htbp]
\centerline{\includegraphics[scale=0.7]{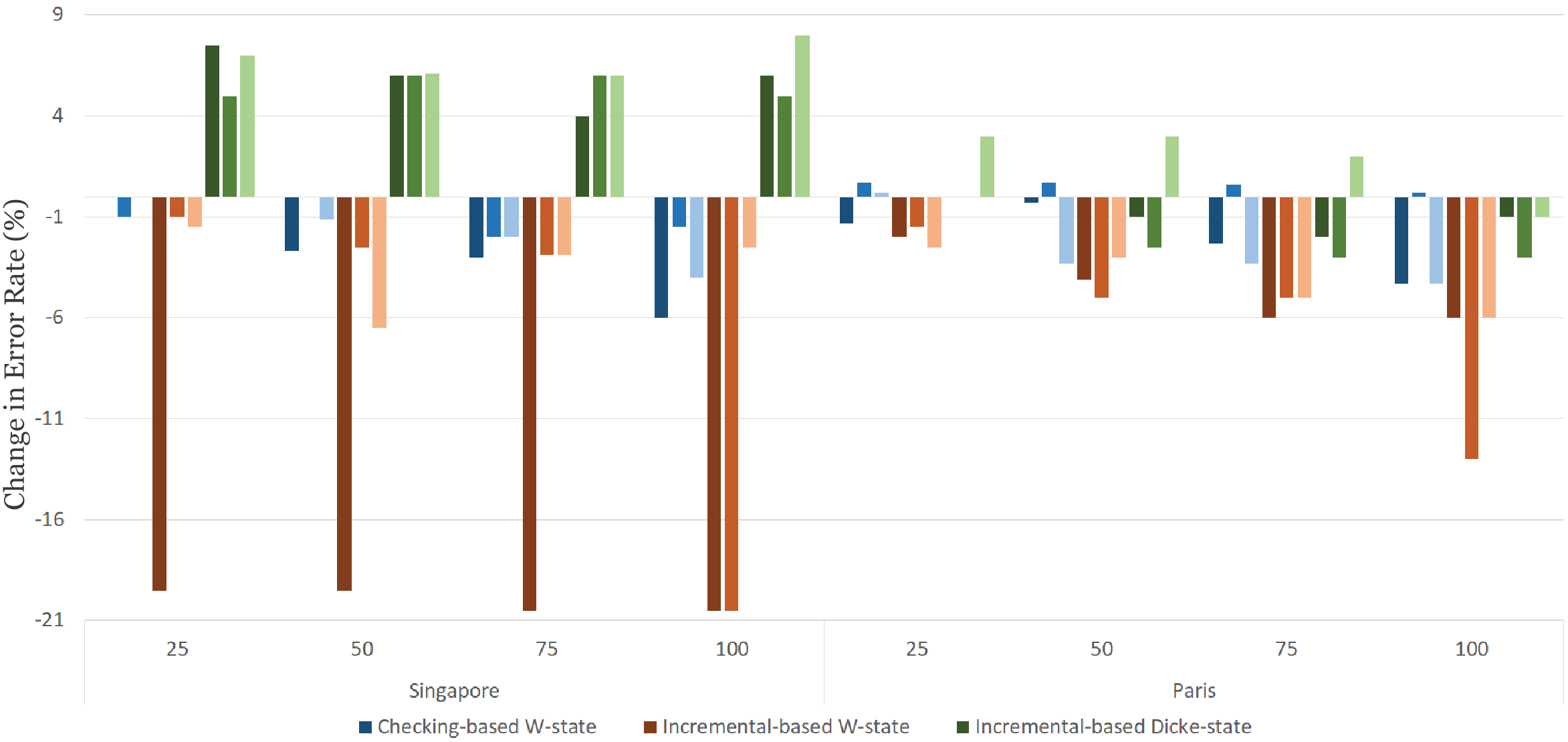}}
\caption{The effect of manipulating the error model of the {\tt ibmq\_singapore} and {\tt ibmq\_paris} devices in decreasing or increasing the error percentage. Each color represents an approach and the shades of the bars represents the modifications type as follows: dark shade  is modifying both $T_1$, $T_2$ and the gate error, the medium shade is modifying $T_1$, $T_2$  and the light shade is modifying the gate error only. $T_1$, $T_2$ are increased by 25, 50, 75, and 100\%, while the gate error is decreased by 25, 50, 75, and 100\%.}
\label{allerror}
\end{figure*}

As can be seen in Fig.~\ref{allerror}, changes in error rate depend on both the implementation of the circuit and the device used for execution. The difference in error rate due to modifications (changing $T_1$, $T_2$, and gate error) can increase the device error up to 7.5\% and decrease down to 20.5\%. It can also be seen that the incremental-based approach with W-state preparation has the most decrease in error rate, especially when modifying both $T_1$, $T_2$, and the gate error. Finally, we can see that changing $T_1$, $T_2$, only leads to better results than modifying the gate error only. 
The reason for modifying the execution parameters caused the error rate to increase in some cases is the fact that the values of $T_1$, $T_2$, or those of gate errors are not constant. They depend on many factors, such as machine maintenance, and the date on which experiments had been conducted.\footnote{Experiments on the real machines to extract these results were conducted from April 15 to April 25, 2020.}

\section{Conclusion}
\label{con}

In this paper, we proposed two approaches to utilizing Grover's Algorithm to solve the $k$-clique problem on a NISQ device, with theoretical asymptotic performance for long-term use. We analyzed the performance of the proposed approaches from different perspectives, such as gate count, gate type, and time complexity, we also analyzed the performance of our method via simulation of six different IBMQ devices. In addition, we showed how theory and implementation could be far apart when it comes to quantum algorithm complexity, due to assumptions about hardware capabilities. Finally, we estimated the closest time our proposed application can be executable with minimal error on a real NISQ device based on the growth chart of quantum processing power introduced by IBM and the current performance of NISQ devices.\\
Future directions for this work will include optimizing both the checking-based and the incremental-based oracles circuits in order to make them executable on NISQ devices sooner than predicted by the QV roadmap. More so, we will work on implementing more gate-efficient oracles to solve the problem with smaller circuit size/depth, as well as extending the algorithm to $k$-clique problem cases where the number of cliques is unknown or larger than one by adding an implementation of a quantum counter to the overall algorithm.
\section*{Acknowledgment}
\label{ack}
This work was supported by MEXT Quantum Leap Flagship Program Grant Number JPMXS0118067285.
The results presented in this paper
were obtained in part using an IBM Q quantum computing system as part of the IBM Q Network. The views expressed are those of the authors and do not reflect the
official policy or position of IBM or the IBM Q team.


\end{document}